\documentclass[12pt]{article}
\usepackage{bbm}

\input epsf
\usepackage{amssymb}
\usepackage[dvips]{graphicx}
\usepackage{amscd}
\usepackage{mathrsfs}
\usepackage{amsmath}

\newtheorem{theorem}{Theorem}[section]

\newcommand{\be}{\begin{equation}}
\newcommand{\ee}{\end{equation}}
\newcommand{\beqa}{\begin{eqnarray}}
\newcommand{\eeqa}{\end{eqnarray}}

\begin{document}
\linespread{1.5} \flushbottom
%\def\be{\begin{equation}}
%\def\ee{\end{equation}}
%nastavenie strany:
\voffset=-10mm %horny okraj
\oddsidemargin=1.0cm %pravy okraj
\textwidth=140mm \textheight=240mm \pagestyle{plain}

\begin{flushright}\today\end{flushright}
\begin{flushright}
LPT-2010-36
\end{flushright}

\begin{center}
{\Large\bf{
Quantum Field Theory on quantized Bergman domain
%On oscillator realization of Conformal Group
%Representations}
}}\vskip1cm
H. Grosse\\{\it Faculty of Physics, University of Vienna,
Boltzmanngasse 5, A-1090 Vienna, Austria }\\
harald.grosse@univie.ac.at
\vskip0.5cm
P. Pre\v{s}najder\\ {\it Faculty of Mathematics, Physics and Informatics,
Comenius University, SK-84248 Bratislava, Slovakia}\\
presnajder@fmph.uniba.sk
\vskip0.5cm Zhituo Wang\\{\it
Laboratoire de Physique Th\'eorique, CNRS UMR 8627,\\
Universit\'e Paris XI,  F-91405 Orsay Cedex, France}\\zhituo.wang@th.u-psud.fr

\end{center}

{\bf Abstract}. We present an oscillator realization of discrete
series representations of group $SU(2,2)$. We give formulas for
the coherent state star-product quantization of a Bergman domain
$D$. A formulation of a (regularized) non-commutative scalar field
on a quantized $D$ is given. \vskip0.5cm
{\bf Subject Classification}.%\vskip1cm
\newpage

\section{Introduction}

The fundamental role of conformal group $SO(4,2)$ for Minkowski space-time was
first stressed by Dirac, \cite{Dirac}. Its covering group $G =
SU(2,2)$ describes conformal properties of spinning particles, see
\cite{Todorov}, where one can found a systematic introduction to
the subject. The group $G$ and its orbits are fundamental for the
twistor theory, \cite{Penrose}. It is also of essential importance
for the ADS-CFT theory\cite{Malda}.

All unitary irreducible representations of the group $G$ were
classified by \cite{Knapp} and \cite{Kihlberg}. More general case of $SU(m,n)$ is
treated in \cite{Klimyk}. The discrete series $SU(2,2)$
representations were used by \cite{Mack} and \cite{Ruhl} for the
investigation of conformal properties of fields on Minkowski
space. The highest/lowest weight of the discrete series of representations has been studied by
\cite{Mack} and \cite{Dobrev}.

The importance of the deformations theory for quantum systems was
first stressed by \cite{Bayen}. The deformation method was
generalized to linear Poisson structures (related to Lie algebras)
in \cite{Rieffel} and to general Poisson structures
in\cite{Konts}. The relation between both approaches was described
in \cite{BenAmar}.

The star-product formula represents an approach going beyond
deformation theory. A general star-product approach, based on
coherent states on co-adjoint orbits \cite{Perelomov}, was
proposed in \cite{GP}, for $SU(2)$ case the star-product formula
was found in \cite{P}, see also \cite{Samann} for $SU(n)$ orbits
(the deformed algebra can be represented as a matrix algebra). A
general formula for compact Lie groups was derived in
\cite{Aleks}.

For non-compact Lie groups the situation is more complicated. The
case $SU(1,1)$ was briefly sketched in \cite{GP}. Similar approach
was applied in \cite{Jakim} to a particular $SU(2,2)$ orbit - the
complex Minkowski space. The corresponding deformed noncommutative
algebra was represented in terms of 4 bosonic oscillator pairs.

A noncommutative field theory may be defined provided the
noncommutative algebra of functions, with some additional
structures, is specified on a configuration space(time). For
$SU(2)$ case this was done in \cite{GP1} and followed by various
other papers. For the noncommutative Heisenberg group, the
formulation of a noncommutative quantum field theory on a Moyal
space, was given in \cite{DFR}. Much more work has been done for
models defined over the Euclidean deformed space-time. This
culminated in \cite{GW} and \cite{Riv}, where a renormalizable
nontrivial 4D model was found and studied. For a recent review see
\cite{Rivasseau}.

The noncommutative space-time model proposed  by H. S. Snyder and
C. N. Yang (see \cite{Snyder}), based on non-compact groups
$SO(4,1)$ and $SO(5,1)$ has not been developed much further,
mainly due to the success of renormalization theory approach to
quantum field theory. In our opinion it could be a right time to
return back to those old ideas.

The paper is organized as follows. In Section 1 we describe the
Lie group $G = SU(2,2)$ and its Lie algebra $\textbf{g} =
su(2,2)$. The relevant mathematical background can be found, e.g.,
in \cite{Kirillov} and \cite{Molchanov}. In Section 3 we present a
simple oscillator realization (in terms of 8 bosonic oscillator
pairs) of most degenerate discrete series of representations which
generalizes more common (Schwinger-Jordan) oscillator realizations
used in the case of compact groups. In Section 4 we construct the
system of coherent states for the representation in question and we give
a corresponding star-product formula for the algebra of functions
on a Bergman domain $D$. Finally, in Section 5 we construct
a quantum field theory model on the quantized Bergman domain $D$.

%\cite{Dirac}, \cite{Todorov}, \cite{Penrose}, \cite{Todorov}\cite{Kirillov},
%\cite{Perelomov}, \cite{Rieffel}, \cite{Klimyk}, \cite{Ruhl},
%\cite{Kihlberg}, \bibitem{Upmeier}

\section{The group $\boldsymbol{SU(2,2)}$ and its Lie algebra}
\subsection{The definition of $\boldsymbol{SU(2,2)}$}
The group $G\,=\,SU(2,2)$ is the subgroup of $SL(4,\bf{C})$
matrices satisfying
%$\mathbb{C}$
\be\label{1.1} g\ =\
\left(\begin{array}{cc}a&b\\c&d\end{array}\right)\, \in\, G\
\Rightarrow\ g^\dagger\,\Gamma\,g\ =\ \Gamma\,,\ \Gamma\ =\
\left(\begin{array}{cc}E&0\\0&-E\end{array}\right)\,.\ee
Here all the entries $a$, $b$, $c$, $d$ are $2\times2$- matrices
and the symbols $0$ and $E$ denote the $2\times2$ zero matrix and
the unit matrix, respectively. Inserting this into (\ref{1.1}) one
obtains two sets of equivalent constraints
\be\label{1.1a} \begin{array}{c} a^\dagger\,a\ =\
E\,+\,c^\dagger\,c\,,\ \ c^\dagger\,d\ =\ E\,+\,b^\dagger\,b\,,\ \
a^\dagger\,b\ =\ c^\dagger\, d\,,\end{array}\ee
or,
\be\label{1.1b} \begin{array}{c} a\,a^\dagger\ =\
E\,+\,b\,b^\dagger\,,\ \ d\,d^\dagger\ =\, E\,+\,c\,c^\dagger\,,\
\ a\,c^\dagger\ =\ b\,d^\dagger .\end{array}\ee
\subsection{Maximal compact subgroup and Bergman domain }
The maximal compact subgroup of $SU(2,2)$ is $K=S(U(2)\times U(2))$ which
consists of the matrices
\begin{equation}
 k\ =\ \begin{pmatrix}
k_1 & 0 \\
0 & k_2 \\
\end{pmatrix},\  k_{1,2}\in U(2),\ \ \  \det(k_1)\,\det(k_2)=1.
\end{equation}
The corresponding Bergman domain $D$ is a kind of Type $1$ Cartan domain
which defined as the group coset space:
\begin{equation}
 D\,=\,G/K.
\end{equation}
It can be represented as the set of all complex $2\times 2$ matrix
\begin{equation}
Z=\begin{pmatrix}
z_{11} & z_{12} \\
z_{21} & z_{22} \\
\end{pmatrix}
\end{equation}
with $Z^\dagger Z<E$. The group action on $D$ is given by:
\begin{equation}\label{action}
 Z'=gZ=(aZ+b)(cZ+d)^{-1}
\end{equation}

The Bergman domain $D$ is a pseudo-convex domain where we could define
Hilbert space $L^2(D, d\mu_N)$ (see (\ref{measure})) of holomorphic functions with reproducing kernel
$K(Z,W)$, where $Z, W\in D$. This reproducing kernel is also called
Bergman kernel and it is well known that
\begin{equation}\label{n4}
 K(Z, W)=\det(E-ZW^{\dagger})^{-N}.
\end{equation}
The Bergman domain $D$ is an $8$ dimensional rank $2$ Hermitian
symmetric space. It is also a K\" ahler manifold of $4$ complex
dimensions, with the metric given by
\begin{equation}
g_{i\bar j}=\frac{\partial^2}{\partial_{z_i}
\partial_{\bar{z_j}}}\log( K(Z,\bar Z))
\end{equation}
The topology of Bergman domain $D$ is nontrivial. It not simply connected but has genus $4$. One way to
calculate the genus is by studying the corresponding complex
Jordan triple. The interested reader could find more details in
(\cite{Upmeier}).

\subsection{The Lie algebra $\textbf{g}$ and the Haar
measure}\label{root} Let $\textbf{g}\,=\,su(2,2)$ be the Lie
algebra of G, so it is real and semisimple. It is formed by
matrices satisfying
\be\label{1.1b0} X^\dagger\,\Gamma\ +\ \Gamma\,X\ =\ 0\ee

Consider the Cartan decomposition of $\textbf{g} $(see
\cite{Molchanov}, \cite{PZh}, \cite{Bob}):
\be\label{phys} \textbf{g}=\textbf{k}+\textbf{p},\ee
where $\textbf{k}$ is the subset of all anti-hermitian matrices in
$\textbf{g}$
\begin{equation}\textbf{k}=\big\{\begin{pmatrix}
A & 0 \\
0 & D \\
\end{pmatrix}: A^\dagger=-A,\,D^\dagger=-D,\,
 \mbox{tr}(A+D)=0,\,
A,D \in M_2(C)\big\}.\end{equation}
The set $\textbf{k}$ is the Lie algebra of the maximal compact
subgroup $K$ in $G$. The subset $\textbf{p}$  of all hermitian
matrices in $\textbf{g}$
\begin{equation}\label{B} \textbf{p} =
\big\{\begin{pmatrix}
0 & B \\
B^\dagger & 0 \\
\end{pmatrix}: B \in M_2(C)\big\}
\end{equation}
is just a linear space and not a Lie algebra.
%So we have:\be\label{phys}\textbf{g}=\big\{X, X\ =\
%\left(\begin{array}{cc}A&0\\ 0&D\end{array}\right)\ +\
%\left(\begin{array}{cc}0&B\\B^\dagger&0\end{array} \right)\,\big\},\ee

Let $\textbf{a}$ be a maximal Abelian subalgebra in $\textbf{p}$ .
We may choose for $\textbf{a}$ the set of all matrices of the form
\begin{equation}
H_\Lambda=\begin{pmatrix}
0 & \Lambda \\
\Lambda & 0 \\
\end{pmatrix}
\end{equation}
where $0$ means the $2\times 2$ matrix with entries zeros and
$\Lambda = \mbox{diag}(\lambda_1,\lambda_2)$ is diagonal $2\times
2$ with $\lambda_1$, $\lambda_2$ real.
%\begin{equation}\Lambda=\begin{pmatrix}\lambda_1 & 0\\ 0 &
%\lambda_2 \\ \end{pmatrix}\end{equation}
The corresponding subgroup consists of all matrices of the type:
\begin{equation}\label{boost}
\delta_\Lambda =\begin{pmatrix}
C & S \\
S & C \\
\end{pmatrix},\ \ \
\begin{array}{c}C=\mbox{diag}(\mbox{ch}\lambda_1,\mbox{ch}
\lambda_2),\\
S=\mbox{diag}(\mbox{sh}\lambda_1,\mbox{sh}\lambda_2).\end{array}
\end{equation}
%where \begin{equation} \label{cartan1} C=\mbox{diag}(\mbox{ch}
%\lambda_1, \mbox{ch} \lambda_2),\ S=\mbox{diag}(\mbox{sh} \lambda_1,
%\mbox{sh} \lambda_2).\end{equation}
Define the dual space $\textbf{a}^\star$ spanned by the elements
$\alpha_i$ satisfying $\alpha_i(H_\Lambda)=\lambda_i$.

Then the roots of $(\textbf{g},\textbf{a})$ are given by
\begin{equation}
\pm 2\alpha_1,\  \pm 2\alpha_2,\  \pm (\alpha_1-\alpha_2)
\end{equation}
with multiplicities $m_{2\alpha_1}=m_{2\alpha_2}=1$ and
$m_{\alpha_1\pm\alpha_2}=2$.

On the root system we choose that the positive Weyl chamber  given
by $C^{+}=\{\lambda_1, \lambda_2\}$ with $\lambda_1>\lambda_2>0$.
Then the positive roots are $2\alpha_1$, $2\alpha_2$ and
$(\alpha_1\pm \alpha_2)$. We use $\Sigma$ donate the set of all
roots and $\Sigma^+$ the set of positive roots.

%The Weyl group $W$ consists of all transformations of the type:
%\begin{equation}
%(\lambda_1, \lambda_2)\mapsto (\epsilon_1 \lambda_{\sigma_1},
%\epsilon_2 \lambda_{\sigma_2})
%\end{equation}
%where $\epsilon_1=\pm 1$, $\epsilon_2=\pm 1$ and $\sigma$ stands for
%permutation of the indices $1$ and $2$.

%Let $\Sigma$ be the set of all roots, and $\Sigma^+$ the set of all
%positive roots.
Define
\begin{equation}
\rho=\frac{1}{2}\sum_{\alpha\in\Sigma^+}m_\alpha \alpha
\end{equation}
So we have
\begin{equation}
\rho=\alpha_1+\alpha_2+(\alpha_1+\alpha_2)+
(\alpha_1-\alpha_2)=3\alpha_1+\alpha_2
\end{equation}

Let $\textbf{a}_c$ be the complex extension of $\textbf{a}$. Follow the same procedure as shown above we could define the complex roots
$\alpha_c\in\textbf{a}_c^*$ as
\begin{equation}\label{harish}
\alpha_c^i(\textbf{a}_c)=\tau_i,\ \  i=1, 2.
\end{equation}
Here $\tau_i$ are complex numbers. The formula of $\rho$ and $\tau_i$ will be used
for constructing the eigenfunction of the radial part of the
invariant Laplacian. See section \ref{field}.

Now we have some physical interpretation of the Lie algebra: the
first summand in formula (\ref{phys}) represents compact
generators of {\it rotations}, whereas the second one represents
non-compact generators {\it boosts}. Since the Lie algebras
$su(2,2)$ and $so(4,2)$ are isomorphic we shall label rotations as
$X_{05}$ and $x_{ab}$, $a,b\,=\,1,2,3,4$, and boosts as $X_{0a}$
and $X_{a5}$ considering them as generators $X_{AB}\,=\,-X_{BA}$,
$A,B\,=\,0,1,\,\dots,\,5$, satisfying $so(4,2)$ commutation
relations
\be\label{1.1b1} [X_{AB},X_{CD}]\ =\ \eta^{AC}\,X_{BD}\,-\,
\eta^{BC}\,X_{AD}\,+\, \eta^{BD}\,X_{AC}\,-\, \eta^{AD}\,X_{BC}\,,\ee
with the metric tensor $\eta^{AB}\,=\,
\mbox{diag}(+1,-1,-1,-1,-1,+1)$. Explicitly, the compact Lie
algebra $\textbf{k}$ is spanned by 7 anti-hermitian matrices and
the basis of $\textbf{p}$ is formed by 8 hermitian matrices given
below:
\[ S_{05} = \frac{i}{2}\left(\begin{array}{cc}1&0\\
0&-1\end{array}\right),\ S_{j4}\ =\ \frac{i}{2}\,\left(
\begin{array}{cc}\sigma_j&0\\0&-\sigma_j\end{array}\right),\
S_{ij} = \frac{i}{2}\varepsilon_{ijk}\left(
\begin{array}{cc}\sigma_k&0\\0&\sigma_k\end{array}\right),\]
$$ S_{k5} = \frac{i}{2}\left(\begin{array}{cc}0&\sigma_k\\
-\sigma_k&0\end{array}\right),\ S_{0k} = \frac{1}{2}
\left(\begin{array}{cc}0&\sigma_k\\\sigma_k&0
\end{array}\right),$$
\be\label{1.7} S_{45} = \frac{1}{2}\left(\begin{array}{cc}0&1
\\1&0\end{array}\right),\ S_{04} = \frac{1}{2}\,\left(
\begin{array}{cc}0&i\\-i&0\end{array}\right),\ee
where $i,j,k\,=\,1,2,3$ and $\sigma_1,\,\sigma_2,\,\sigma_3$ are
usual Pauli matrices. \vskip0.5cm

%The maximal compact subgroup $K\,=\,S(U(2)\times U(2))$, with the
%Lie algebra  is formed by block-diagonal matrices $g$. Its Lie
%algebra $Lie(K)\,=\,s(u(2)\oplus u(2))$ is formed by matrices
%\be\label{1.1b2} X\ =\ \left(\begin{array}{cc}A&0\\
%0&D\end{array}\right),\ \ \  A^\dagger\,=\,-A,\ D^\dagger\,=\,-D,\
%\mbox{tr}(A+D)\,=\,0.\ee \vskip0.5cm

The principal Cartan subalgebra
\be\label{1.1b3} \textbf{h}\ =\ \textbf{a}\,\oplus\,\textbf{u}\ee
of $G$ is spanned by three commuting generators: two noncompact
$X_{45}$ and $X_{03}$ span $\textbf{a}$, and the additional
compact one $X_{12}$ spans $\textbf{u}$. The corresponding
subgroups we shall denote as $H$, $A$ and $U$:  $H\,=\,A\times U$.

%is a principal Cartan subgroup with Lie algebra $Lie(H)$ formed by
%one compact generator and two non-compact generators:
%\be\label{1.1b3} X\,=\,\left(\begin{array}{cc}A&0\\ 0&A\end{array}\right)
%\,+\,\left(\begin{array}{cc}0&B\\ B&0\end{array}\right),\ \ \
%\begin{array}{c}A\,=\,i\mbox{diag}(\varphi,-\varphi)\\
%B\,=\,\mbox{diag}(\lambda_1,\lambda_2)\end{array}\ee
%In usual notations the Lie algebra $Lie(H)$ has the basis $X_{45}$,
%$X_{03}$ and $X_{12}$.

Any element of $G$ possesses a unique Cartan decompositions
\be\label{1.1b4} g\ =\ k\,\delta\,\tilde{q}\ =\ \tilde{k}\,\delta\,q\,,\ee
where $\delta$ is some pure positive non-compact element of $H$ with
positive $\lambda_1$ and $\lambda_2$ given by formula (\ref{boost}).
%
%\be\label{1.1b5} \delta\ =\ \left(\begin{array}{cc}C&S\\ S&C\end{array}\right),\ \ \ \begin{array}{c} C\,=\,\mbox{diag}(\mbox{ch}\,\lambda_1,\mbox{ch}\,
%\lambda_2)\\ S\,=\,\mbox{diag}(\mbox{sh}\,\lambda_1,\mbox{sh}\,\lambda_2)\end{array}\ee
%
%with positive $\lambda_1$ and $\lambda_2$.
Further, $k\,=\,\tilde{k}\,u$, and $q\,=\,u\,\tilde{q}$ are
elements of $K$, and $u$ is the element of a compact subgroup $U$
in $H$.

The Haar measure $dg$ on $G\,=\,SU(2,2)$ is, in the parametrization
(\ref{1.1b4}), given as
\be\label{1.4b} dg\ \equiv\ dg(\lambda,k,q)\ =\ \rho(\lambda_1,
\lambda_2)\,d\lambda_1\,d\lambda_2\,d\tilde{k}\,d\tilde{q}\,du\,,\ee
where $d\tilde{k}$ and $d\tilde{q}$ denotes the normalized
measures on $K/U$, and $du$ is the usual measure on $U$ (see
\cite{Ruhl}). The explicit form of $\rho(\lambda_1, \lambda_2)$ is
constructed from the positive roots (see Section \ref{root}):
\begin{equation}
\rho(\Lambda)=\prod_{\alpha\in\Sigma^+}|\sinh\alpha(T)|^{m_\alpha}
\end{equation}
where $m_\alpha$ is the multiplicity of the positive roots. So we
have:
% \frac{\pi^2}{8}\,
\begin{equation} \rho(\lambda_1,\lambda_2)\ =\
\mbox{sh}^2\,(\lambda_1+\lambda_2)
\,\mbox{sh}^2\,(\lambda_1-\lambda_2)\,\mbox{sh}\,(2\lambda_1)\,
\mbox{sh}\,(2\lambda_2)\,.
\end{equation}
\vskip1cm

\section{Discrete series representation of $\boldsymbol{SU(2,2)}$}

The group $G\,=\,SU(2,2)$ possesses principal, supplementary and
discrete series of unitary irreducible representations, see e.g.,
\cite{Kihlberg}, \cite{Klimyk}. The discrete series of
representations is given by:
\begin{equation}\label{rep}
 T_g f(Z)=[\det(C Z+ D)]^{-N}f(Z'), \ N=4, 5, 6\cdots
\end{equation}
where $Z'$ is given by formula (\ref{action}) and 
$f(Z)\in L^2(D,d\mu_N)$ with the measure
\be\label{measure} d\mu_N(\bar{Z}, Z)\ =\ 
c_N\,d\bar{Z}\,dZ\,\mbox{det}
\,(E-Z^\dagger Z)^{N-4}. \ee
The normalization constant $c_N\,=\,\pi^{-4}(N-1)(N-2)^2 
(N-3)$ guarantees that the function $f_0(Z)\,\equiv 1$ 
has a unit norm, see \cite{Ruhl}.
%where $f(Z)\in L^2(D)$ and $Z'$ is given by formula
%(\ref{action}). Below we construct an oscillator realization of
%most degenerate discrete series representations depending on one
%natural number $N$.

We introduce a $4\times 2$ matrix $\hat{Z} = (\hat{z}_{a\alpha})$,
$a\,=\,1,\,\dots,\,4$, $\alpha\,=\,1,2$, of bosonic oscillators
acting in Fock space and satisfying commutation relations
$$ [\hat{z}_{a\alpha},\hat{z}^\dagger_{b\beta}]\ =\ -\Gamma_{ab}\,
\delta_{\alpha\beta}\,, $$
\be\label{2.1}  [\hat{z}_{a\alpha},\hat{z}_{b\beta}]\ =\
[\hat{z}^\dagger_{a\alpha},\hat{z}^\dagger_{b\beta}]\ =\ 0\,,\ee
where $\Gamma$ is a $4\times 4$ matrix defined in (\ref{1.1}). It can
be easily seen that for all  $g\,\in\,SU(2,2)$ these commutation
relations are invariant under transformations:
\be\label{2.1a} \hat{Z}\ \mapsto\ g\,\hat{Z},\ \ \ \hat{Z}^\dagger\
\mapsto\ \hat{Z}^\dagger\,g^\dagger .\ee
Since, $\Gamma\,=\,\mbox{diag}(+1,+1,-1,-1)$ the upper two rows in
$\hat{Z}$ corresponds to creation operators whereas the lower ones
to annihilation operators:
\be\label{2.2}  \hat{Z}\ =\ \left(\begin{array}{c}\hat{a}^\dagger\\
\hat{b}\end{array}\right):\ \   [\hat{a}_{\alpha\beta},
\hat{a}^\dagger_{\gamma\delta}]\ =\  [\hat{b}_{\alpha\beta},
\hat{b}^\dagger_{\gamma\delta}]\ =\ \delta_{\alpha\beta}\,
\delta_{\gamma\delta},\ \  \alpha,\beta,\gamma,\delta\,=\,1,2. \ee
and all other commutation relations among oscillator operators vanish.
The Fock space  ${\cal F}$ in question is generated from a normalized
vacuum state $|0\rangle$, satisfying  $\hat{a}_{\alpha\beta}\,
|0\rangle\,=\,\hat{b}_{\alpha\beta}\,|0\rangle\,=\,0$, by repeated
actions of creation operators:
\be\label{2.3} |m_{\alpha\beta},\,n_{\alpha\beta}\rangle\ =\
\prod_{\alpha\beta}\,\frac{(\hat{a}^\dagger_{\alpha\beta}
)^{m_{\alpha\beta}}\,(\hat{b}^\dagger_{\alpha\beta}
)^{n_{\alpha\beta}}}{\sqrt{m_{\alpha\beta}!\,n_{\alpha\beta}!}}
\,|0\rangle\,.\ee
We shall use the terminology that the state $|m_{\alpha\beta},
\,n_{\alpha\beta}\rangle$ contains $m=\sum m_{\alpha\beta}$
particles $a$ and $n=\sum n_{\alpha\beta}$ particles
$b$.\vskip0.5cm

The Lie algebra $su(2,2)=so(4,2)$ acting in Fock space can be
realized in terms of oscillators. To any $4\times 4$ matrix
$X\,=\,(X_{ab})$ we assign the operator
\be\label{2.3a} \hat{X}\,=\,-\mbox{tr}(\hat{Z}^\dagger\Gamma
X\hat{Z})\ =\ -\hat{z}^\dagger_{a\alpha}\,
\Gamma_{ab}\,X_{bc}\,\hat{z}_{c\alpha} \,,\ee
with $\hat{Z}^\dagger$ and $\hat{Z}$ given in (\ref{2.1}) and
(\ref{2.2}) in terms of oscillators. Using commutation relations
for annihilation and creation operators the commutator of
operators $\hat{X}\,=\,-\mbox{tr}(\hat{Z}^\dagger\Gamma X\hat{Z})$
and $\hat{Y}\,=-\,\mbox{tr}(\hat{Z}^\dagger\Gamma Y\hat{Z})$ can
be easily calculated:
\be\label{2.4} [\hat{X},\hat{Y}]\ =\ [\mbox{tr}
(\hat{Z}^\dagger\Gamma X\hat{Z}),\mbox{tr}(\hat{Z}^\dagger\Gamma
Y\hat{Z})]\ =\ -\mbox{tr}[\hat{ Z}^\dagger\,\Gamma\,(XY-YX)\,
\hat{Z}]\,.\ee
%%%%%%%%%%%%%%%%%%%%%%%%%%%%%%%%%
It can be easily seen that the {\it anti-hermitian operators}
\be\label{2.3b} \hat{X}_{AB}\,=\,-\mbox{tr}(\hat{Z}^\dagger\Gamma
X_{AB}\hat{Z}),\ \ A,B=0,1, \dots,5,\ee
%\hat{z}^\dagger_{a\alpha}\,\Gamma_{ab}\,X_{bc}\,\hat{z}_{c\alpha}
with $X_{AB}$ given in (\ref{1.7}), satisfy in Fock space the
$su(2,2)\cong so(4,2)$ commutation relations (\ref{1.1b1}).  The
assignment
\be\label{2.4a}  g\,=\,e^{\xi^{AB}X_{AB}}\,\in\,SU(2,2)
\Rightarrow\ \hat{T}(g)\,=\,e^{\xi^{AB}\hat{X}_{AB}} \ee
then defines a {\it unitary $SU(2,2)$ representation} in Fock
space.

The adjoint action of $\hat{T}(g)$ on operators reproduces
(\ref{2.1a}). In terms of $a$ and $b$-oscillators in block-matrix
notation this can be rewritten as
\be\label{2.4b}  g=\left(\begin{array}{cc}a&b\\c&d\end{array}
\right):\ \begin{array}{cc} \hat{T}(g)\,\hat{a}^\dagger\,
\hat{T}^\dagger(g)\,=\,a\,\hat{a}^\dagger+b\,\hat{b}, &
\hat{T}(g)\,\hat{a}\hat{T}^\dagger(g)\,=\,
\hat{a}\,a^\dagger+\hat{b}^\dagger\,b^\dagger,\\
\hat{T}(g)\,\hat{b} \,\hat{T}^\dagger(g)\,=\,
d\,\hat{b}+c\,\hat{a}^\dagger, & \hat{T}(g)\,\hat{b}^\dagger\,
\hat{T}^\dagger(g)\,=\, \hat{b}^\dagger\,d^\dagger+
\hat{a}\,c^\dagger.\end{array}\ee
Since any $g\,\in\,SU(2,2)$ possesses Cartan decomposition
(\ref{1.1b4}) we shall discuss separately rotations and special
boosts given in (9). For rotations we obtain a mixing of
annihilation and creation operators of a same type:
\be\label{k} k=\left(\begin{array}{cc} k'&0\\0&k^{\prime\prime}
\end{array}\right):\ \begin{array}{cc} \hat{T}(k)\,\hat{a}^\dagger
\,\hat{T}^\dagger(k)\,=\,k'\,\hat{a}^\dagger, & \hat{T}(k)\,
\hat{a}\hat{T}^\dagger(g)\,=\,\hat{a}\,k^{\prime\dagger},\\
\hat{T}(g)\,\hat{b}\,\hat{T}^\dagger(k)\,=\,k^{\prime\prime}\,\hat{b},
&\hat{T}(k)\,\hat{b}^\dagger\,\hat{T}^\dagger(k)\,=\,\hat{b}^\dagger
\,k^{\prime\prime\dagger}.\end{array} \ee
However, for special boosts from $\textbf{a}$ we obtain Bogolyubov
transformations:
\be \delta =\left(\begin{array}{cc}C&S\\S&C\end{array} \right):\
\begin{array}{cc} \hat{T}(\delta)\,\hat{a}^\dagger\,\hat{T}^\dagger(g)
\,=\,C\,\hat{a}^\dagger+S\,\hat{b}, & \hat{T}(\delta)\,\hat{a}
\hat{T}^\dagger(\delta)\,=\,\hat{a}\,C+\hat{b}\,S,\\
\hat{T}(\delta)\,\hat{b}\,\hat{T}^\dagger(\delta )\,=\,C\,
\hat{b}+S\,\hat{a}^\dagger, &\hat{T}(\delta )\,\hat{b}^\dagger\,
\hat{T}^\dagger(\delta )\,=\,\hat{b}^\dagger\,C+\hat{a}\,S,
\end{array}\ee
%%%%%%%%%%%%%%%%%%%%%%%%%%%%%%%%%
with $C$ and $S$ determined in (\ref{boost}).

Using the explicit form of matrices $X_{AB}$, following from
(\ref{1.1b0}), the action of generators can be described in terms
creation and annihilation of $a$- and $b$-particles:

(i) The action of {\it rotation} generators results in a replacement
of some $a$-particle by an other $a$-particle and by replacement of
$b$-particle by other $(ab)$-particle.

(ii)  The action of {\it boost} generators results in creation of a
pair $(ab)$ of particles or in a destruction of  $ab$  pair.

In this context it is useful to consider lowering and rising
operators labeled by arbitrary $2\times 2$ complex\ matrix
$B\,=\,(B_{\beta\gamma})$ entering (\ref{B}), that annihilate and
create $ab$ pairs:
\be\label{2.5} \hat{T}^B_-\ =\
\hat{a}_{\alpha\beta}\,B_{\beta\gamma}\,\hat{b}_{\gamma\alpha},\ \
\ \hat{T}^B_+\ =\ (\hat{T}^B_+)^\dagger\ =\
\hat{a}^\dagger_{\alpha\beta}\,B^*_{\beta\gamma}\,
\hat{b}^\dagger_{\gamma\alpha}.\ee
More specifically, we can consider $B = E_{\beta \gamma}$ (the
matrix with $1$ in the intersection of $\beta$-th row and
$\gamma$-th column and $0$'s otherwise). The corresponding
lowering and rising operators are:
\be\label{2.5a} \hat{T}^{\beta \gamma}_-\ =\
\hat{a}_{\alpha\beta}\,\hat{b}_{\gamma\alpha},\ \ \ \hat{T}^{\beta
\gamma}_+\ =\ (\hat{T}^{\beta \gamma}_+)^\dagger\ =\
\hat{a}^\dagger_{\alpha\beta}\,\hat{b}^\dagger_{\gamma\alpha}.\ee
In (\ref{2.5}) and (\ref{2.5a}) the summation over $\alpha$ is
understood. Any boost can be uniquely expressed as complex
combinations of operators $T^{\beta \gamma}_-$ and $T^{\beta
\gamma}_+$. It follows from (\ref{2.5}) that the operator
\be\label{2.6}   \hat{N}\equiv
\frac{1}{2}(\hat{N}_{\hat{b}}-\hat{N}_{\hat{b}})\ =\
\frac{1}{2}(\hat{b}^\dagger_{\alpha\beta}\hat{b}_{\alpha\beta}
\,-\,\hat{a}^\dagger_{\alpha\beta}\hat{a}_{\alpha\beta})\ =\
\frac{1}{2}Tr(\hat{Z}^\dagger\Gamma\hat{Z})\,-\,2 \ee
commutes with all generators $\hat{X}_{AB}$, $A,B\,=\,0,1,\,
\dots,\,5$. \vskip0.5cm

Below, we shall restrict ourselves to most degenerate discrete
series representations which are specified by the eigenvalue
of the operator $ \hat{N}$  in the representation subspace.
We start to construct the representation space ${\cal F}_N$
from a distinguished normalized state containing lowest number
of particles:
%c^{-1}_N\,=\,N!\,\sqrt{N+1},
\begin{eqnarray}\label{2.7} |x_0\rangle\ &=&\
\frac{\mbox{det}(\hat{b}^\dagger)^{N-1}}{(N-1)!\,\sqrt{N}}\ |0\rangle,\ N=1, 2, \cdots
\nonumber
\\
 &=&\ \frac{1}{\sqrt{N}}\sum_{n=0}^{N-1}
(-1)^n\,|\,0,0,0,0\,;\,N-1-n,n,n,N-1-n\rangle\,.\end{eqnarray}
Here, $\hat{b}^\dagger\,=\,(\hat{b}^\dagger_{\alpha\beta})$ is
$2\times 2$ matrix of $b$-particle creation operators, and $N$ is
a natural number that specifies the representation: $
\hat{N}\,|x_0\rangle\,=\,(N-1)\,|x_0\rangle$. All other states in the
representation space are obtained by the action of rising
operators given in (\ref{2.5}): such states contain besides $(2N-2)$
$b$-particles a finite number of $ab$ pairs. \vskip0.5cm

The maximal compact subgroup $K\,=\,S(U(2)\times U(2))$ is the
stability group of the state $|x_0\rangle$. The group action for
$k\,=\,\mbox{diag}(k',k^{\prime\prime})$ reduces just to the phase
transformation  (see (\ref{k})):
\be\label{2.7a} \hat{T}(k)\,|x_0\rangle\ =\
\mbox{det}(k^{\prime\prime\dagger})^{N-1}\,\mbox{det}(k')\,
|x_0\rangle\ =\
\mbox{det}(k^{\prime\prime})^{-N}\,|x_0\rangle\,.\ee
The first factor comes from $\hat{b}^\dagger\,\mapsto\,
\hat{b}^\dagger\,k^{\prime\prime\dagger}$ (see (\ref{k})), whereas
the second factor comes from
$\hat{T}(k)\,|0\rangle\,=\,\mbox{det}(k')\,|0\rangle$ (due to the
anti-normal ordering of the compact generators containing
$\hat{a}$ and $\hat{a}^\dagger$).

Let us calculate  the mean values of the operator $\hat{T}(g)$
in the state $|x_0\rangle$: $\omega_0(g) = \langle x_0|\,
\hat{T}(g)\,|x_0\rangle$.  Using Cartan decomposition $g = k\,
\delta\,q$  and the action(\ref{2.7}) of rotations we obtain:
$$ \omega_0(g)\ =\ \langle x_0|\,\hat{T}(g)\,|x_0\rangle\ =\
\langle x_0|\,\hat{T}(k)\,\hat{T}(\delta)\,\hat{T}(q)\,|x_0
\rangle $$
\be\label{2.7b} =\ \mbox{det}(k^{\prime\prime})^{-N}\,
\mbox{det}(q^{\prime\prime})^{-N}\, \langle
x_0|\,\hat{T}(\delta)\,|x_0\rangle .\ee
Thus it is enough to calculate the mean value for the special
boost:
\be\label{2.8}  \delta\,=\,\left(\begin{array}{cc}C&S\\
S&C\end{array}\right) = \left(\begin{array}{cc}E&0\\
T&E\end{array}\right)\left(\begin{array}{cc}C&0\\
0&C^{-1}\end{array}\right)\left(\begin{array}{cc}E&T\\
0&E\end{array}\right)\,=\,t_+\,t_0\,t_-.\ee
 Here $C=\mbox{diag}(\mbox{ch}\,\lambda_1,\mbox{ch}\,\lambda_2)$,
 $S=\mbox{diag}(\mbox{sh}\,\lambda_1,\mbox{sh}\,\lambda_2)$ and
 $T=\mbox{diag}(\mbox{th}\,\lambda_1,\mbox{th}\,\lambda_2)$.
 In the representation in question, the matrices $t_+$ and $t_-$
 are exponents of rising and lowering operators respectively:
%$T=SC^{-1}$
$$ \hat{T}(t_+)\,=\, e^{\mbox{tr}(\hat{b}^\dagger\,T\,
\hat{a}^\dagger)},\ \ \hat{T}(t_-)\,=\, e^{-\mbox{tr}(a\,T\,b)}.
$$
Since $ \hat{T}(t_-)$  contains  $\hat{a}$ its action does not
affect  $|x_0\rangle$, and similarly $\hat{T}(t_+)$  containing
$\hat{a}^\dagger$ does not affect $\langle x_0|$. The only
non-trivial action comes from
%%%%%%%%%%%%%%%%%%%%%%%%%%%%%%%%%%%%%%%%%%%%%%
\be\label{2.9} \hat{T}(t_0)\,=\, 
e^{-\,\mbox{tr}(\hat{a}\,\Lambda\,\hat{a}^\dagger)\,-\,
\mbox{tr}(\hat{b}^\dagger\,
\Lambda\,\hat{b})},\ \ \ 
\Lambda\,=\,\ln C\,.\ee
Consequently,
\be\label{2.9a} \hat{T}(t_0)|x_0\rangle
\ =\ \mbox{det}\,(C)^{-N}\,|x_0\rangle\,. \ee
The last equality follows from the identity
$$  e^{-\mbox{tr}(\hat{b}^\dagger\,\Lambda\,b)}\,
\mbox{det}\,(\hat{b}^\dagger)^N\ =\ e^{-N\,\mbox{tr}(\Lambda)}\,
\mbox{det}(\hat{b}^\dagger)^N\, e^{-\mbox{tr}(\hat{b}^\dagger\,
\Lambda\,b)} $$
%%%%%%%%%%%%%%%%%%%%%%%%%%%%%%%%%%%%%%%%%%%%%
(which can be proven, e.g. by induction in $N$). From equations
(\ref{2.7a}) and  (\ref{2.9}) we obtain remarkably simple
results:
\be\label{2.10} \omega_0(g)\ =\ \langle
x_0|\,\hat{T}(g)\,|x_0\rangle\ =\
\mbox{det}(k^{\prime\prime})^{-N}\,\mbox{det}
(C)^{-N}\,\mbox{det}(q^{\prime\prime})^{-N}
 =\ \det(d)^{-N}.\ee
Here $d\,=\,k^{\prime\prime}\,C\,q^{\prime\prime}$ is the
lower-right block of matrix $g$ (see the Cartan decomposition in
(\ref{2.7b})). We recovered the results valid in the holomorphic
representation (\ref{rep}). 
\section{The star product}

Starting from the normalized state $|x_0\rangle\,\in \,{\cal F}_N$
we shall construct the Perelomov's system of coherent states for
the representation in question. We choose a set of boosts of the
form:
\be\label{2.11} g_x =  k\,\delta\,k^\dagger\ =\
\left(\begin{array}{cc}k'\,C\,k^{\prime\dagger}&k'\,
S\,k^{\prime\prime\dagger}\\ k^{\prime\prime}\,S\,
k'&k^{\prime\prime}\,C\,k^{\prime\prime\dagger}\end{array}\right)
\equiv \left(\begin{array}{cc}C' &\tilde{S}\\ \tilde{S}^\dagger
& C^{\prime\prime}\end{array}\right) \in G/K,\ee
where $k\equiv\tilde{k}= \mbox{diag}(k',k^{\prime\prime})$ is an
element of $K$ with eliminated Cartan $U(1)$ factor,
$C=\mbox{diag}(\mbox{ch\,}\lambda_1,\mbox{ch\,}\lambda_2)$ and
$S=\mbox{diag}(\mbox{sh\,}\lambda_1,\mbox{sh\,}\lambda_2)$. Thus $ x
= x(k,\delta)$ depends on 8 parameters: 6 of them, hidden in $k$,
are compact, and there are 2 non-compact ones $\lambda_1$ and
$\lambda_2$ representing special boosts. To any boost $g_x$, given
in (\ref{2.11}), we assign coherent coherent state (see,
\cite{Perelomov})
\be\label{2.12} |x\rangle\ =\  \hat{T}(g_x)\,|x_0\rangle\ =\
\hat{T}(k\,\delta\,k^\dagger)\,|x_0\rangle,\ \ x = x(k,\delta).
\ee

{\it Note}: Any $ x = x(k,\delta),$  can be uniquely assigned to
the $2\times 2$ complex matrix $z =
k'\,T\,k^{\prime\prime\dagger}$,
$T=\mbox{diag}(\mbox{th\,}\lambda_1,\mbox{th\,} \lambda_2)$,
forming the bounded Bergman domain $D\cong G/K$.

Let us consider operators in the representation space of the form
\be\label{2.13} \hat{F}\ =\ \int_G dg\,\tilde{F}(g)\,
\hat{T}(g),\ee
where $\tilde{F}(g)$ is a distribution on a group $G$ with
compact support $\mbox{supp}\,\tilde{F}$. To any such operator
we assign function on $G/K$ by the prescription
\be\label{2.14} F(x)\ =\ \langle x|\,\hat{F}\,|x\rangle\ =\
\int_G dg\,\tilde{F}(g)\,\omega(g,x),\ee
where
\be\label{2.14a} \omega(g,x)\ \equiv\ \langle x|\,\hat{T}(g)
\,|x\rangle\ =\  \omega_0(g^{-1}_x\,g\,g_x).\ee
This equation combined with (\ref{2.10}) offers an explicit
form of $\omega(g,x)$ and is well suited for calculations.

The star-product of two functions $ F(x)= \langle x|\,\hat{F}
\,|x\rangle$ and $ G(x)= \langle x|\,\hat{G}\,|x\rangle$ was
defined in \cite{GP}:
$$ (F\star G)(x)\ =\ \langle x|\,\hat{F}\hat{G}\,|x\rangle\
=\ \int_{G\times G} dg_1 dg_2\,\tilde{F}(g_1)\,\tilde{G}(g_2)\,
\omega(g_1g_2,x)$$
\be\label{2.15} =\ \int_G dg\,(\tilde{F}\circ\tilde{G})(g)\,
\omega(g,x).\ee
In (\ref{2.15}) the symbol $\tilde{F}\circ\tilde{G}$ denotes the
convolution in the group algebra $\tilde{\cal A}_G$ of compact
distributions:
\be\label{2.16} (\tilde{F}\circ\tilde{G})(g)\ =\ \int_G dh\,
\tilde{F}(gh^{-1})\,\tilde{G})(h).\ee
Obviously, the mapping $\tilde{F}\,\mapsto\,F$ given in
(\ref{2.14}) is a homomorphism of the group algebra $\tilde{\cal
A}_G$ into the star-algebra ${\cal A}^\star_G$ of functions
(\ref{2.14}) on $D=G/K$. The quantized Bergman domain $D_\star$ we
identify with the noncommutative algebra of functions ${\cal
A}^\star_G$ on $D$. \vskip0.5cm

{\it Note}: We point out that as in the case of usual
distributions, the convolution product may exist even for
distribution with non-compact support provided there are satisfied
specific restrictions at infinity. \vskip0.5cm

It can be easily seen that
$$ \mbox{supp}\,(\tilde{F}\circ\tilde{G})\,\subset\,(\mbox{supp}
\,\tilde{F})\,(\mbox{supp}\,\tilde{G})\,
%$$ $$
\equiv\,\{ g=g_1 g_2\,|\,g_1\in\mbox{supp}\,\tilde{F},\,g_2\in
\mbox{supp}\,\tilde{G}\}. $$
Consequently, for a non-compact group there are two classes
of group algebras:

(i) The first one is generated by distributions with a general
compact support and the corresponding group algebra is simply
the full algebra $\tilde{\cal A}_G$ defined in (\ref{2.15}).

(ii) The second one is formed by distributions $\tilde{F}$ with
$\mbox{supp}\,\tilde{F}$ subset of a subgroup $H\subset K$, form a
sub-algebra $\tilde{\cal A}_H$ of  the group algebra $\tilde{\cal
A}_G$. \vskip0.5cm

In the second class there are are two interesting extremal cases:

(a) $\tilde{\cal A}_{\{e\}}$ corresponding to the trivial subgroup
$H = \{e\}$ in $G = SU(2,2)$ ($\tilde{\cal A}_{\{e\}}$ is
isomorphic to the enveloping algebra ${\cal U}(su(2,2))$, see
e.g., \cite{Kirillov} or \cite{Molchanov}).

(b) $\tilde{\cal A}_K$ corresponding to the maximal compact
subgroup $K$ in $G$. \vskip0.5cm
%Below we shall describe star-algebras  ${\cal A}^\star_{\{e\}}$
%and ${\cal A}^\star_K$, and finally, we shall discuss the full
%star-algebra ${\cal A}^\star_G$. \vskip0.5cm
%{\it The algebra ${\cal A}^\star_{\{e\}}$}.

The deformation quantization on Lie group co-orbits in terms of
the Lie group convolution algebra  $\tilde{\cal A}_{\{e\}}$  was
introduced by \cite{Rieffel}. Here we follow the related coherent
state construction of the star-star  on Lie group co-orbits
proposed in \cite{GP}.

Any distribution  $\tilde{F}$ can be given as a linear combination
of finite derivatives of the group $\delta$-function, i.e., as a
linear combination of distributions
\be\label{2.17}  \tilde{F}_{A_1B_1\dots A_nB_n}(g)\ =\
({\cal X}_{A_1B_1}\,\dots\, {\cal X}_{A_nB_n}\delta)(g),\ee
where ${\cal X}_{AB}$ is the left-invariant vector field
representing the generator $X_{AB}$. The explicit form of ${\cal X}_{AB}$ in terms of the coordinates of the Bergman domain has been given by \cite{Esteve} and \cite{barut}.

Inserting this into (\ref{2.14}) we obtain the corresponding function from
${\cal A}^\star_{\{e\}}$
\be\label{2.18}  F_{A_1B_1\dots A_nB_n}(x)\ =\ (-1)^n
({\cal X}_{A_nB_n}\,\dots\, {\cal X}_{A_1B_1}\omega)(g,x)|_{g=e}.\ee
Here we used the fact that the operators ${\cal X}_{AB}$ are
anti-hermitian differential operators with respect to the group
measure $dg$. From (\ref{2.14}) it follows directly that
$$ (F_{A_1B_1\dots A_nB_n}\star F_{C_1D_1\dots C_mD_m})(x) $$
\be\label{2.19}   =\ (-1)^{n+m}({\cal X}_{A_nB_n}\,\dots\,
{\cal X}_{A_1B_1}\,{\cal X}_{C_mD_m}\,\dots\,
{\cal X}_{C_1D_1}\omega)(g,x)|_{g=e}.\ee
Equations (\ref{2.18}) and (\ref{2.19}) describe explicitly the
homomorphism ${\cal U}(su(2,2))\to{\cal A}^\star_{\{e\}}$.
\vskip0.5cm

Using exponential parametrization of the group element $g =
e^{\xi^{AB}X_{AB}}$ formula for the symmetrized function
(\ref{2.18}) takes simple form:
$$ F_{\{A_1B_1\dots A_nB_n\}}(x)\ =\ (-1)^n
(\partial_{\xi_{A_1B_1}}\,\dots\, \partial_{\xi_{A_nB_n}}
\omega)( e^{\xi^{AB}X_{AB}},x)|_{\xi=0} $$
\be\label{2.20} =\ (-1)^n\,\langle x|\,\hat{X}_{\{A_1B_1}\,\dots\,
\hat{X}_{A_nB_n\}}\\|x\rangle, \ee
where $\{\,\dots\,\}$ means symmetrization of double indexes and
$\xi=0$ means the evaluation at $\xi_{AB}=0$ for
$A,B=0,1,\dots,5$. Symmetrized form a basis of the algebra in
question and symmetrized elements from the center of algebra
correspond to Casimir operators. In the series of representation
in question all Casimir operators are given in terms of a single
operator $\hat{N}$ given in (\ref{2.6}) which is represented by a
constant function $N(x) = \langle x|\hat{N}|x\rangle = N$.
\vskip0.5cm

{\it Example 1.: The function $\omega(g,x)$}. Let us calculate the
function $\omega(g,x) = \omega_0(g_x^{-1}\,g\,g_x)$ explicitly.
Taking $g$ and $g_x$ as in (\ref{1.1}) and (\ref{2.11}) we have to
calculate the product of three matrices:
 $$ g_x^{-1}\,g\,g_x\ =\ \left(\begin{array}{cc}C' & -\tilde{S}\\
-\tilde{S}^\dagger & C^{\prime\prime}\end{array}\right) 
\left(\begin{array}{cc}a&b\\ c&d\end{array}\right) 
\left(\begin{array}{cc}C' &\tilde{S}\\
\tilde{S}^\dagger & C^{\prime\prime}\end{array}\right)\ \equiv\
\left(\begin{array}{cc}a_x&b_x\\ c_x&d_x\end{array}\right),$$
where $d_x\,=\,C^{\prime\prime}\,d\,C^{\prime\prime}
\,+\,C^{\prime\prime}\,c\,\tilde{S}^\dagger\,-\,
\tilde{S}\,c\,C^{\prime\prime}\,-\,\tilde{S}^\dagger\,a\,
\tilde{S}$. Using equation (\ref{2.10}) for $\omega_0(g)$ we obtain
\begin{eqnarray}\label{2.20b} &&\omega(g,x)=\det(d_x)^{-N}\nonumber\\
&=&\det(E-z_x^\dagger z_x)^{N}\det(d\,+\,cz^\dagger_x\,-\,z_xb\,-\,z_xaz^\dagger_x)^{-N}
\end{eqnarray}
Here $z_x = C'^{-1}\tilde{S} = k'Tk^{\prime\prime\dagger}$ is
$2\times 2$ complex matrix from the Bergman domain. In this 
form the expression is convenient for calculations.\vskip0.5cm

{\it Example 2.: Fock space realization of the co-adjoint orbit}:
Our aim is to calculate the coordinates
\be\label{2.21} \xi_{AB}(x) = \frac{1}{N}\,\langle
x|\hat{X}_{AB}|x\rangle = \frac{1}{N}\,\langle x_0|\hat{T}^\dagger
(g_x)\,\hat{X}_{AB}\hat{T}(g_x)|x_0\rangle, \ee
for $A,B\,=\,0,1,\,\dots,\,5$. Taking into account (\ref{2.4b}) we
see that $\xi_{AB}(x) = D^{CD}_{AB}(g_x)\,\xi_{AB}(x_0)$, where
$(D^{CD}_{AB}(g))= Ad^*_g$ is the matrix corresponding to the
group action in co-adjoint algebra. Therefore it is sufficient to
evaluate the coordinates at $x_0$: $\xi_{AB}(x_0) =
\frac{1}{N}\,\langle x_0|\hat{X}_{AB}|x_0\rangle$. A simple
calculation gives: $\xi_{45}(x_0) = 1$ with all other
$\xi_{AB}(x_0) = 0$. We see that $\xi_{AB}(x)$ just forms the
co-adjoint orbit generated from $\xi_{AB}(x_0)$. \vskip0.5cm

{\it Example 3. The star-product of coordinates $\xi_{AB}(x)$}: We
have
$$ (\xi_{AB}\star\xi_{CD})(x)\ =\ \frac{1}{2N^2}\,\langle
x|\{\hat{X}_{AB},\hat{X}_{CD}\}|x\rangle\ +\ \frac{1}{2N^2}\,
\langle x|[\hat{X}_{AB},\hat{X}_{CD}]|x\rangle ,$$
where $\{\,\dots\,\}$ denotes anti-commutator and $[\,\dots\,]$ is
commutator. Therefore, the second term is
$$ \frac{1}{2N^2}\,\langle x|[\hat{X}_{AB},\hat{X}_{CD}]|x
\rangle\ =\ \frac{1}{2N}\,f^{EF}_{AB,CD}\,\xi_{EF}(x), $$
where we used the definition of $\hat{X}_{AB}$ and the short-hand
notation for the commutator (\ref{1.1b1}): $[X_{AB},X_{CD}] =
f^{EF}_{AB,CD} X_{EF}$. The first term is proportional to the
symmetrized function $F_{\{AB,CD\}}$ and we can use (\ref{2.20}):
$$ \frac{1}{2N^2}\,\langle x|\{\hat{X}_{AB},\hat{X}_{CD}\}
|x\rangle\ =\ (1 + A_N)\,\xi_{AB}(x)\,\xi_{CD}(x)\,+\,B_N\,
\delta_{AB,CD}, $$
where we have a usual point-wise product of functions in the first
term and $\delta_{AB,CD} = (1/2)(\delta_{AC}\delta_{BD}\,
-\,\delta_{AD}\delta_{BC})$ in the second one. The coefficients
$A_N$ and $B_N$ are of order $1/N$. Last two equations give
\be\label{2.22} (\xi_{AB}\star\xi_{CD})(x)\ =\ (1+ A_N)\,
\xi_{AB}(x)\,\xi_{CD}(x) \,+\,\frac{1}{2N}\,f^{EF}_{AB,CD}
\,\xi_{EF}(x)\,+\,B_N\, \delta_{AB,CD}. \ee
We see that the parameter of the non-commutativity is $\lambda_N =
1/N$. For $N\,\to\,\infty$ we recover the commutative product.

\begin{theorem}
 The star product (\ref{2.15}, \ref{2.22}) is associative and
 invariant under the transformation of $SU(2,2)$ group.
\end{theorem} %\begin{proof}
The proof of this theorem follows directly the definition of the
star product. %\end{proof}

%{\it The algebra ${\cal A}^\star_K$}. \vskip0.5cm

%{\it The algebra ${\cal A}^\star_G$}. \vskip0.5cm
\section{Quantum field on a Bergman domain
$\boldsymbol{D}$}\label{field}
\subsection{The invariant Laplacian on $\boldsymbol{D}$}
The invariant Laplacian $\Delta_N$ is defined by:
\begin{equation}
 \Delta_N \hat {T_g}=\hat{ T_g}\Delta_N
\end{equation}
where $\hat{T_g}$ is the representation operator given by
(\ref{2.4a}). We have (see \cite{ Hua}, \cite{PZh}):
\begin{eqnarray}
&&\Delta_N=tr[(E-ZZ^{\dagger})\bar{\partial}_Z\cdot
(E-Z^{\dagger}Z)\cdot \partial'_Z]\\
&+&\det(E-Z^{\dagger}
Z)^{-N}tr[(E-ZZ^{\dagger})\bar{\partial}_Z\cdot
(E-Z^{\dagger}Z)\cdot
\partial'_Z(\det(E-Z^{\dagger}Z)^N)]\nonumber
\end{eqnarray}
where $\partial_Z = (\partial_{z_{ij}})$ is the $2\times 2$ matrix
of differential operators in the variables $Z=(z_{ij})\in D$,
$\bar{\partial}_Z$ and $\partial'_Z$ denote, respectively the
complex conjugate and the transpose of the matrix operator
$\partial_Z$. It is understood that the operators
$\bar{\partial}_Z$ and $\partial'_Z$ do not differentiate the
matrices $E-ZZ^{\dagger}$ and $E-Z^{\dagger}Z$. 

The Laplacian
$\Delta_N$ is self-adjoint on $L^2(D,d\mu_N)$ with respect to the measure
given by (\ref{measure}).

In what follows we consider only the radial part of the invariant
Laplacian $\Delta_N$, as this part that contains information about
most interesting physical quantities, e.g., the energy levels. The
radial part of the invariant Laplacian could be constructed from
the roots system introduced in section \ref{root}, (see
(\cite{PZh}, \cite{Bob}) for more details). The radial part of the
invariant Laplacian reads:
\begin{equation}
\Delta^r_N=\omega^{-1}(\sum_{i=1}^2\frac{1}{4}L_i\,-\,
\frac{N}{2}\mbox{th}\,\lambda_i\,\partial_{\lambda_i})\omega
\end{equation}
where
\begin{equation}
\omega=2(\mbox{ch}\,2\lambda_1\,-\,\mbox{ch}\,2\lambda_2)
\end{equation}
and
\begin{equation}
L_i=\frac{\partial^2}{\partial{\lambda_i}^2}+2\mbox{cth}
\,2\lambda_i\,\partial_{\lambda_i}
\end{equation}

Let $\Phi(N,\tau_1, \tau_2)$ be the eigenfunction of $\Delta^r_N$,
we have (see\cite{PZh}):
\begin{equation}
\Delta^r_N\Phi(N, \tau_1,\tau_2)=-\frac{1}{4}[2(N-1)^2
+\tau_1^2+\tau_2^2]\Phi(N,\tau_1,\tau_2)
\end{equation}
where $\tau_1$ and $\tau_2$ are given by (\ref{harish}).

The operator $-\Delta_N$ is positive: it has the continuous
spectrum
\begin{equation}
\big[-\frac{1}{2}(N-1)^2,+\infty\big)
\end{equation}
for arbitrary $\tau_i$, and the discrete finite spectrum
%$\begin{pmatrix} \big[\frac{n-1}{2}\big]+1\\2\\ \end{pmatrix}$
\begin{equation}
%\frac{1}{4}[2(N-1)^2-(N-1-2l_1)^2-(N-1-2l_2)^2]\,=\,
(N-1)(l_1+l_2)\,+\,l_1^2+l_2^2
\end{equation}
for $\tau_j$, $\ j=1,2$, imaginary:
\begin{equation} \tau_j=-i(N-1-2l_j),\
l_j=0, 1,\cdots,\big[\frac{N-1}{2}\big].
    \end{equation}
Here $\big[\frac{N-1}{2}\big]$ means the integer part of
$\frac{N-1}{2}$. The discrete spectrum consists of $\frac{1}{2}\
k(k-1)$ points, where $k=\big[\frac{N-1}{2}\big]$.
%So for fixed $N$ the discrete spectrum is finite.
%for \begin{equation}\tau_j=-i(n-1-2l_j),\ j=1,2\ .\end{equation}

\subsection{A quantum field theory model}
We shall present a construction of a real scalar (Euclidean) field
theory on a quantized Bergman domain. The action of the model in
question reads:
%\mathscr{L}
\begin{equation}\label{model}
S[\Phi]\ =\ \int d\mu_N(\xi)\ [-\frac{1}{2}\Phi\star\Delta_N\Phi+
\frac{1}{2}m^2 \Phi\star\Phi+V_\star(\Phi)]\,,
\end{equation}
where $\Phi=\Phi(\xi)$ is a real scalar field depending on the
noncommutative coordinates $\xi$ and $d\mu_N(\xi)$ is the measure
(\ref{measure}) expressed in terms of $\xi$.
%For simplicity we suppose that $\Phi$ is a polynomial of $\xi(x)$.
We suppose that $V$ is a polynomial of $\Phi$ bounded from below.
Then we expand the scalar field $\Phi$ in terms of the
eigenfunction of the invariant Laplacian:
\begin{equation}
\Phi=\int d\tau_1 d\tau_2\,C_N(\tau_1,\tau_2)\Phi(N,
\tau_1,\tau_2)+\sum_{l_1,l_2} C_{N,l_1,l_2}\Phi(N, l_1,l_2),
%\Phi=\int d\tau_1 d\tau_2\sum_n C_n(\tau_1, \tau_2)\Phi(n,
%\tau_1,\tau_2)+\sum_{l_1,l_2,n} C_{n, l_1, l_2}\Phi(n, l_1,l_2)
\end{equation}
where we integrated over the continuous part of spectrum and
summed up over the discrete part of the spectrum. The coefficients
$C_N(\tau_1, \tau_2)$ and $C_{N, l_1, l_2}$ are arbitrary real
numbers.

%Remark that the Lagrangian introduced in (\ref{model}) is
%conformal invariant.
The quantum mean value of some polynomial field functional
$F[\Phi]$ is defined as the functional integral over fields
$\Phi$:
\begin{equation}
 \langle F[\Phi]\rangle\ =\ \frac{\int D\Phi\,e^{-S[\Phi]}\,
 F[\Phi]}{\int D\Phi\,e^{-S[\Phi]}},
\end{equation}
where $D\Phi=D_x d\Phi(x)\cong\prod_{\tau_j} dC_N(\tau_j)$ and
$S[\Phi]$ denotes the corresponding action (\ref{model}).

%%%%%%%%%%%%%%%%%%%%%%%%%%%%%%%%%%%%%%
For the free field propagator we recover the quantum field theory
results on a commutative Bergman domain $D$
\begin{equation}
<\Phi, \Phi>=\frac{1}{m^2+\frac{1}{4}[2(N-1)^2+\tau_1^2+\tau^2_2]}
\end{equation}
which is valid for arbitrary $\tau_j$:
%For different values of $\tau_j$ the theory has different behavior
%of divergence.
\begin{itemize}
 \item For the discrete part of the spectrum where
 $\tau_j=-i(N-1-2l_j)$, the quantum field theory is finite, it
 possesses a cutoff at the maximal energy level $N$.
\item For the continuous part of the spectrum where the parameters
$\tau_j$ are arbitrary, the theory is divergent. But it could be
made finite after proper renormalization. This point will be
studied in more detail in future publications.
\item When $N=\rm{finite}$, we have $\frac{1}{N}$ corrections for
the vertices coming from the lowest order of the star product. The
divergent behavior is similar to the semiclassical case.
\end{itemize}

\section{Concluding remarks}
In this paper we introduced an oscillator realization of the
discrete series of $SU(2,2)$ representations. We performed a
deformation quantization over the corresponding coset space
$D=SU(2,2)/S(U(2)\times U(2))$. We presented an explicit
expression of the star-product over $D$. Using this star product
we constructed a QFT model over this noncommutative Bergman domain
$D_\star$. This method can be applied to other $SU(m,n)$ type I
Cartan domains, see \cite{PGW}, where $SU(2,1)$ is discussed in
detail. Such results are of interest for both physics and
mathematics. From the physical point of view, $SU(2,2)$ is the
maximal symmetry group of the (compactified) Minkowski space. It
is also of interest for the ADS-CFT correspondence, as $SU(2,2)$
is the double cover of $SO(4,2)$ conformal group. As the $SU(2,2)$
module, the Bergman domain $D$ is a K\" ahler manifold which is
important in supersymmetric Quantum field theory and string
theory. In addition the Bergman domain $D$ has nontrivial Shilov
boundary, and the quantization of $D$ could help us to understand
this boundary problem in the framework of noncommutative geometry.
These aspects are under study and will be discussed in a
forthcoming paper. \newpage%\vskip1cm

{\bf Acknowledgements} \vskip0.5cm The author ZT Wang is very
grateful to the Physics department of University of Vienna and the
Erwin-Schr\" odinger Institute  for hospitality and financial
support. The work of P Pre\v snajder was supported by the project
VEGA 1/100809 of the Slovak Ministry of Education .

\end{document}